\documentclass[twocolumn,fleqn]{article} \usepackage{amsmath,amssymb,amsfonts}
\usepackage{graphicx} 			
\usepackage{url}
\usepackage{siunitx} 
\title {Analysis of COVID-19 infection waves in Japan by Avrami equation} 
\author {Yoshihiko Takase \thanks {Chiba JICA Senior Volunteers Association}} 

\begin{document} 
\maketitle 

\abstract {}
Last time, an attempt was made to analyze the first wave of COVID-19 in Japan from February to May 2020 by the Avrami equation. This time, all infection waves that occurred repeatedly from February 2020 to September 2021 were simulated by developing the last work. The entire waveform was basically simulated by the superposition of five major waves, and the least-squares method was applied to each wave to determine the parameters of the equation. Despite the simulation using the single equation assuming that the parameters were constants, the accuracy of the simulated cumulative infection $D$ was as good as $95\% CI / D < 2.5\%$ after the second wave when $D$ exceeded 10,000. Since the simulation was highly accurate, it was effective in predicting the number of infections in the near future and in considering the cause of changes in the number of infections observed as the detailed structure of the waveform.

\section {Introduction}

In the previous work\cite{covid19_ng_model}, we attempted to analyze the first wave of COVID-19 in Japan from February to May, 2020 by the Avrami (or JMAK) equation\cite{avrami} that describes the phase transformation dynamics and is another approach than the conventional SIR model and its expanded models\cite{sir_model1,sir_model2,sir_model3,sir_model4,sir_model5}. As its entity model, the physical process of the linear growth of nuclei forming randomly in the parent phase was assumed. This process was quantitatively determined by three parameters, the initial susceptible $D_\mathrm{s}$, the domain growth rate $K$, and the nucleation decay constant $\nu$. 

This time, the target duration of the analysis is expanded to include all major waves as well as small initial waves from February 2020 to September 2021, and detailed analysis will be performed using the same theory.

The purpose of this study is to simulate the entire waveform of the recurring infection waves by developing the last work.

\section {Data analysis} 
\subsection {Theoretical basis}

In the previous work\cite{covid19_ng_model}, two cases where the domain grew one-dimensionally and two-dimensionally from randomly generated nuclei were compared. This time, each major wave was represented by a superposition of multiple waves without ignoring small waves. As a result, the one-dimensional growth model was not considered because the two-dimensional growth model provided more physically reasonable parameter values.

The number of time-dependent daily new infections $J(t)$ (referred to as the daily infections later) is\cite{covid19_ng_model} 

\begin {equation} 
\label {eq:avrami_j} 
  J(t) = D_\mathrm{s} \left( -\frac{K^2}{\nu} f_1 \right) \exp\left(-\frac{K^2}{\nu^2} f_2 \right), 
\end {equation} 
where $t$ is the time,  $D_\mathrm{s}$ is the initial susceptible, $K^2 = 2\pi G^2 l_c N_0$, $G$ the growth speed, $l_c$ the domain thickness, $N_0$ the number of active points for nucleus, $\nu$ the decay constant, $f_1 = 1 - \mathrm{e}^{-\nu t} - \nu t$ and $f_2 = 1 - \mathrm{e}^{-\nu t} - \nu t + (\nu t)^2 / 2$.

The number of time-dependent cumulative daily new infections $D(t)$ (referred to as the total infections later) is expressed by the following equation\cite{covid19_ng_model},

\begin {equation} 
\label {eq:avrami_d} 
  D(t) = D_\mathrm{s} \left[ 1- \exp \left ( -\frac{K^2} {\nu^2} f_2 \right ) \right ]. 
\end {equation} 

\subsection {Result of the total wave analysis}

\begin {figure} [p] 
	\centering 
	\includegraphics[width=8.0cm]{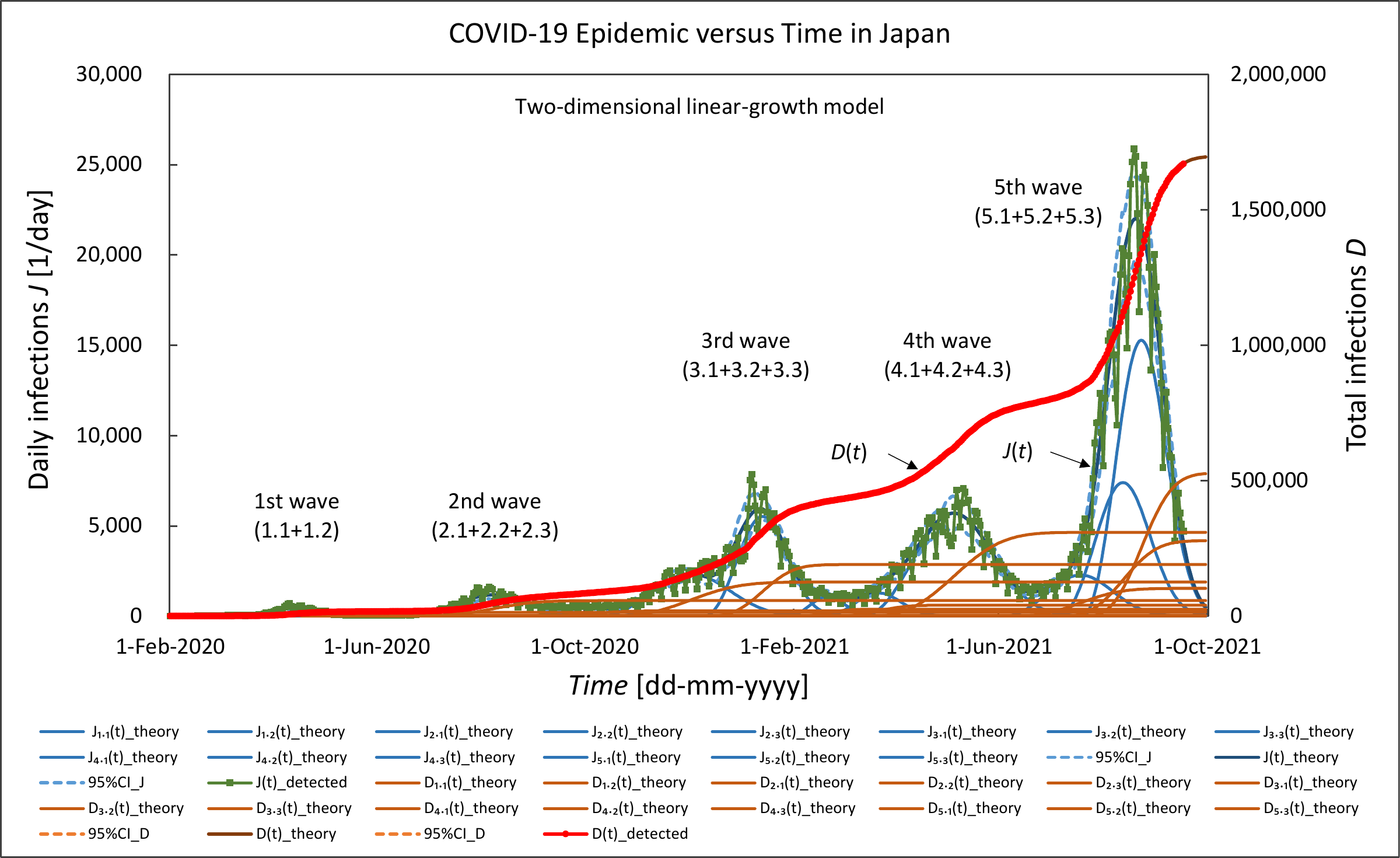} 
	\caption {Detected and simulated results of COVID-19 infection waves over the entire duration in Japan. 
	The dark green and the red markers are the detected daily and total infections, respectively. The blue curve is the theoretical value of the daily infections for each wave and the dark blue curve is the sum of them, $J(t)$. The brown curve is the theoretical value of the total infections for each wave and the dark brown curve is the sum of them, $D(t)$. The broken line is the $95\% CI$.
	} 
	\label {fig:allwaves_linear} 
\end {figure}

\begin {figure} [p] 
	\centering 
	\includegraphics[width=8.0cm]{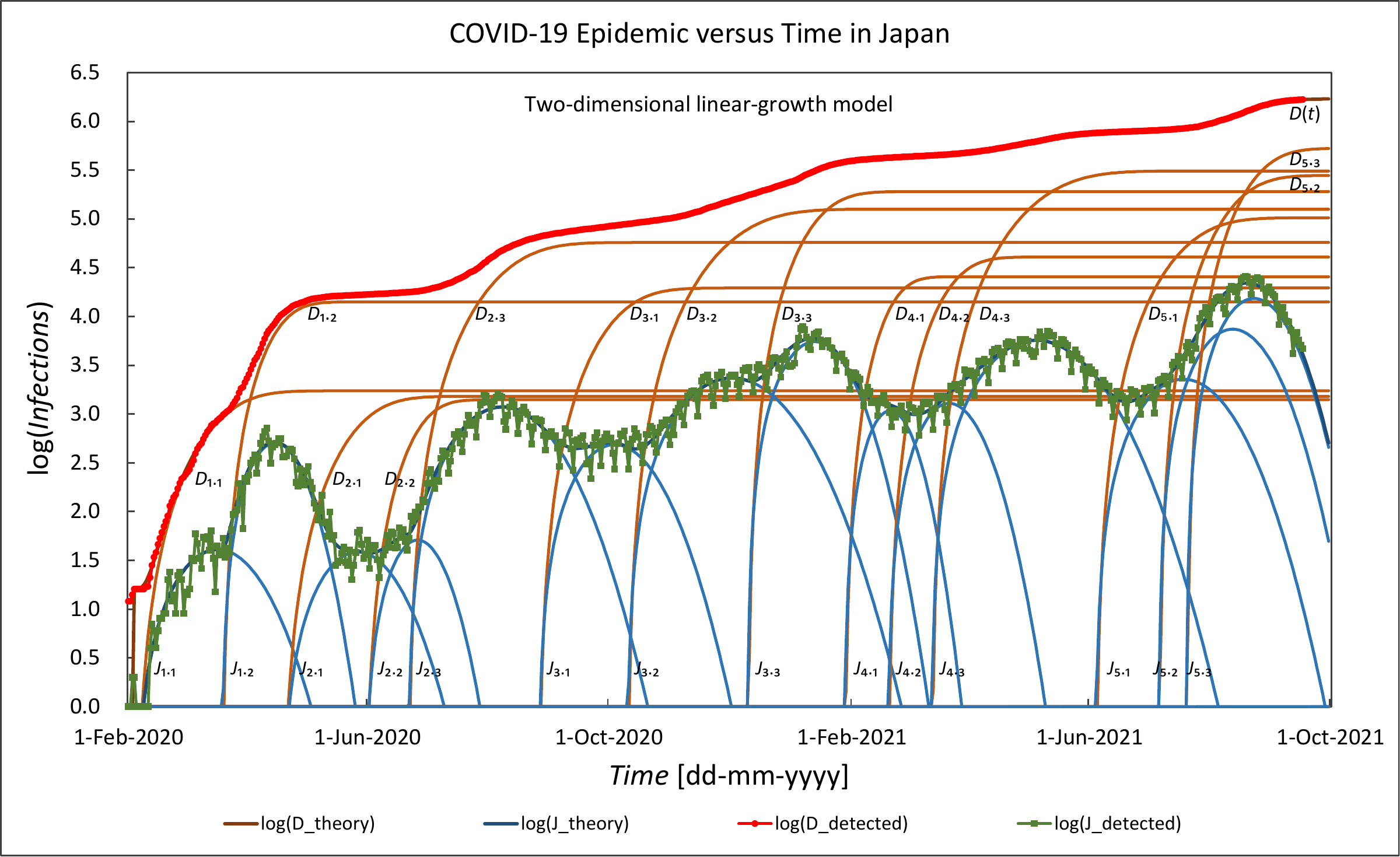} 
	\caption {Detected and simulated results of COVID-19 infection waves over the entire duration in Japan. 
	The meanings of the markers and the curves are the same as those in Fig. \ref{fig:allwaves_linear}. The entire waveform was simulated with a total of 14 wave superpositions grouped into 5 major waves.
	}
	\label {fig:allwaves_semilog} 
\end {figure}

First, the results of analysis of the entire waveform of the daily and the total infections are shown in Figs. \ref{fig:allwaves_linear} and \ref{fig:allwaves_semilog} as linear and semi-logarithmic plots, respectively. 
The dark green and the red markers are the detected daily and total infections, respectively. The blue curve is the theoretical value of the daily infections for each wave and the dark blue curve is the sum of them, $J(t)$. The brown curve is the theoretical value of the total infections for each wave and the dark brown curve is the sum of them, $D(t)$. The broken line is the $95\% CI$ calculated from the moving standard deviation with 7-days sliding window.

The entire waveform was simulated with a total of 14 wave superpositions grouped into 5 major waves. The detected value of the daily infections fluctuates remarkably in a weekly cycle after the second wave. This is because Sunday is a holiday in Japan and many medical institutions are also closed. The detected value of the daily infections $J(t)$ fluctuates in this way, but the detected total infections $D(t)$ shows a smooth curve. The least-squares method was applied to the curve of the total infections to determine the three parameters. The accuracy of the simulated total infections $D$ was as good as $95\% CI / D < 2.5\%$ after the second wave when $D$ exceeded 10,000.

\subsection {Superposition analysis method} 

In order to make the theoretical value more consistent with the detected value than the last work\cite{covid19_ng_model}, the first wave was expressed as the superposition of the initial 1.1th wave and the main 1.2th wave.

First, the least-squares method is applied to the root mean squate (RMS) between detected and theoretical values of the time-dependent characteristic of the 1.1th wave total infections to determine the three parameters, $D_\mathrm{s}$, $K^2$ and $\nu$.

Next, as shown in Eqs. \ref{eq:superpos_j} and \ref{eq:superpos_d}, the value obtained by subtracting the theoretical value of the previously obtained wave from the detected value is regarded as the new detected value of the wave to be analyzed. The least-squares method is applied to RMS between new detected and theoretical values to determine the three parameters of the wave to be analyzed. This operation is repeated from the 1.2th wave to the 5.3th wave.

The new detected value of the daily infections $J_\mathrm{newdet} (t)$ to be analyzed is

\begin {equation} 
	\label {eq:superpos_j} 
	J_\mathrm{newdet} (t) = J_\mathrm{det} (t) - \sum_{i=1}^{W_1} \sum_{j=1}^{W_2} J_{i.j} (t), 
\end {equation} 
where $J_\mathrm{det} (t)$ is the detected value of the daily infections, $W_1$ is the id number of each wave (1, 2, 3, 4, or 5), and $W_2$ is the sub-id number of each wave (1, 2, or 3). For example, if $W_1 = 5$ then $W_2 = 2$ (if $i = 1$) or 3 (if $ i = 2, 3, 4,$ and 5).

The new detected value of the total infections $D_\mathrm{newdet} (t)$ to be analyzed is

\begin {equation} 
	\label {eq:superpos_d} 
	D_\mathrm{newdet} (t) = D_\mathrm{det} (t) - \sum_{i=1}^{W_1} \sum_{j=1}^{W_2} D_{i.j} (t), 
\end {equation} 
where $D_\mathrm{det} (t)$ is the detected value of the total infections, $W_1$ and $W_2$ are the same as those of Eq. \ref{eq:superpos_j}.

The duration of each wave to which the least-squares method is applied is shown in Table \ref{table:waves_duration}.

\begin{table} 
\caption{The duration of each wave to which the least-squares method is applied.} 
	\label{table:waves_duration} 
	\centering
	\begin{tabular} {ccc} 
	\hline 
	Wave & Start day & End day \\ 
	\hline \hline 
	1.1th & 04-Feb-2020 & 18-Mar-2020 \\ 
	1.2th & 19-Mar-2020 & 07-May-2020 \\ 
	\hline 
	2.1th & 18-Apr-2020 & 07-Jun-2020 \\ 
	2.2th & 30-May-2020 & 22-Jun-2020 \\ 
	2.3th & 21-Jun-2020 & 08-Sep-2020 \\ 
	\hline 
	3.1th & 26-Aug-2020 & 13-Oct-2020 \\ 
	3.2th & 10-Oct-2020 & 20-Dec-2020 \\ 
	3.3th & 10-Dec-2020 & 02-Feb-2021 \\ 
	\hline 
	4.1th & 28-Jan-2021 & 28-Feb-2021 \\ 
	4.2th & 19-Feb-2021 & 22-Mar-2021 \\ 
	4.3th & 13-Mar-2021 & 13-Jun-2021 \\ 
	\hline 
	5.1th & 04-Jun-2021 & 11-Jul-2021 \\ 
	5.2th & 06-Jul-2021 & 22-Jul-2021 \\ 
	5.3th & 20-Jul-2021 & 17-Sep-2021 \\ 
	\hline 
	\end{tabular} 
\end{table}

\begin {figure} 
	\centering 
	\includegraphics[width=8.0cm] {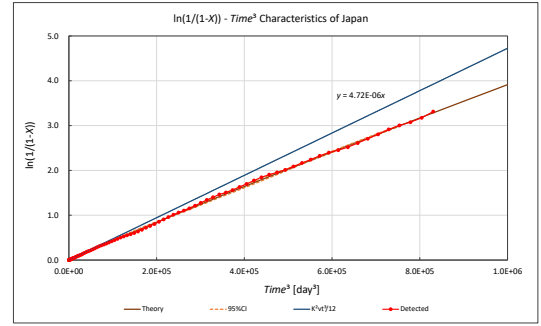} 
	\caption {The $\ln (1/(1-X)) - Time^3$ plot of the
	Avrami equation for the 4.3th wave. $X$ is the fraction of domain that has been
	transformed by formation and growth of nuclei. The red marker is the detected
	value, the dark brown line represents the theory, the dark blue straight line
	is the approximate straight line when $\nu t$ is small enough, and the broken
	line represents the $95\% CI$.} 
	\label {fig:y_t3_4_3} 
\end {figure}

As an example, the $\ln (1/(1-X)) - Time^3$ plot of the Avrami equation\cite{leo_mandelkern} for the 4.3th wave is shown in Fig. \ref{fig:y_t3_4_3}, where $X$ is the fraction of domain that has been transformed by formation and growth of nuclei. The red marker is the detected value, the dark brown line is the theoretical value, the dark blue straight line is the approximate straight line when $\nu t$ is small enough, and the broken line is the $95\% CI$ calculated from the moving standard deviation with the sliding window set to 7 days. The characteristic shows linearity with a small deviation.

\begin {figure} 
	\centering 
	\includegraphics[width=8.0cm] {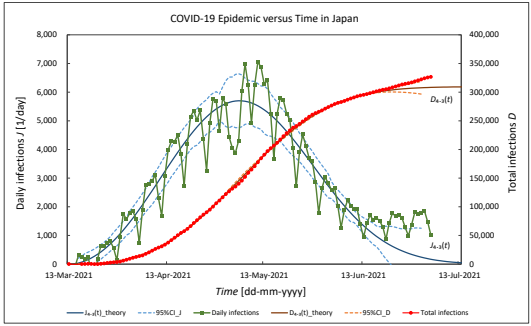} 
	\caption {The daily and the total infections versus time characteristics of the 4.3th wave. 
	The dark green and the red markers are the detected values of the daily and the total infections, respectively.
	The blue and the brown curves represent the theory of the daily and the total infections,
	respectively. The broken line represents the $95\% CI$.} 
	\label {fig:inf_t_4_3} 
\end {figure}

Figure \ref{fig:inf_t_4_3} shows the time-dependent characteristics of the daily and the total infections of the 4.3th wave. The meanings of the markers and the curves are the same as those in Fig. \ref{fig:allwaves_linear}. The theoretical curve for the total infections closely approximates the detected value. The detected value of the daily infections fluctuates greatly in a weekly cycle, but is approximated by the theoretical curve when evaluated in the $95\% CI$.

\subsection {Determined parameters} 

Table \ref{table:three_parameters} shows the parameters determined for each of the 14 waves, $D_\mathrm{s}$, $K^2$, $\nu$, and $t_\mathrm{on}$, where $t_\mathrm{on}$ is the rise time defined as the time taken by the $D(t) - t$ characteristic to change from the 10\% value to the 90\% value.
   
\begin{table} 
  \caption {Parameters determined for each of the 14 waves} 
  \label {table:three_parameters} 
  \centering 
  \begin {tabular} {crrrr} 
	\hline 
	Wave & \multicolumn{1}{c}{$D_\mathrm{s}$} & \multicolumn{1}{c}{$K^2$} & \multicolumn{1}{c}{$\nu$} & \multicolumn{1}{c}{$t_\mathrm{on}$}\\ 
	& [person] & [$1/\si{day}^2$] & \multicolumn{1}{c}{[$1/\si{day}$]} & \multicolumn{1}{c}{[$\si{day}$]} \\ 
	\hline \hline 
	1.1th & 1,729 & 0.00846 & 0.00660 & 41 \\ 
	1.2th & 14,100 & 0.01200 & 0.01800 & 28 \\ 
	\hline 
	2.1th & 1,517 & 0.00797 & 0.00900 & 40 \\ 
	2.2th & 1,401 & 0.01882 & 0.01100 & 28 \\ 
	2.3th & 57,300 & 0.00623 & 0.00600 & 49 \\ 
	\hline 
	3.1th & 19,569 & 0.00498 & 0.01500 & 41 \\ 
	3.2th & 125,770 & 0.00432 & 0.00650 & 54 \\ 
	3.3th & 190,417 & 0.01063 & 0.01000 & 34 \\ 
	\hline 
	4.1th & 25,570 & 0.01893 & 0.01800 & 24 \\ 
	4.2th & 40,647 & 0.01097 & 0.01300 & 32 \\ 
	4.3th & 309,360 & 0.00506 & 0.00800 & 54 \\ 
	\hline
	5.1th & 102,542 & 0.00600 & 0.00800 & 45 \\ 
	5.2th & 278,797 & 0.01000 & 0.00800 & 37 \\ 
	5.3th & 528,277 & 0.01093 & 0.00960 & 34 \\
	\hline
  \end {tabular} 
\end {table}

\begin {figure} 
	\centering 
	\includegraphics[width=8.0cm] {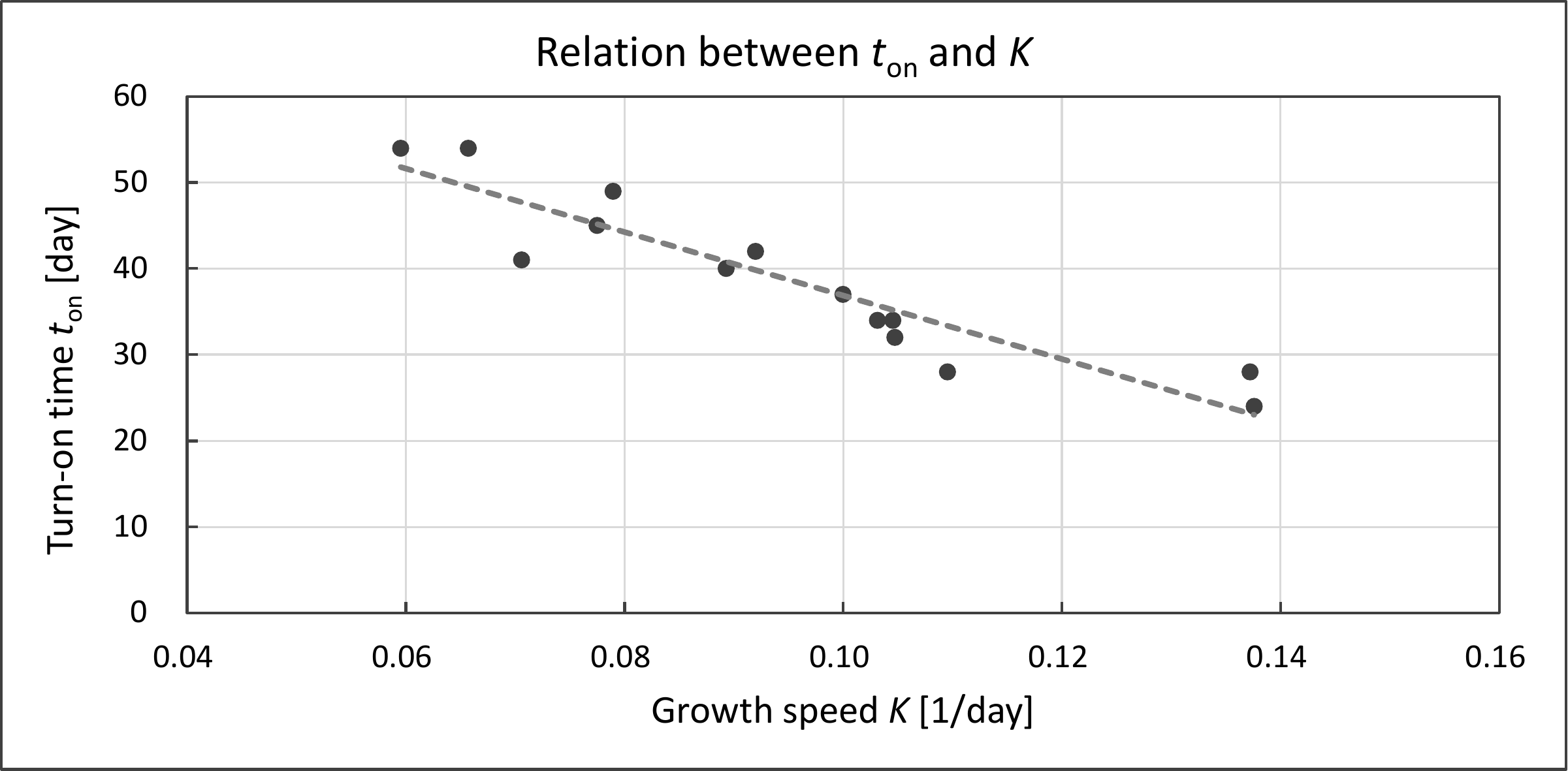} 
	\caption {Relation between the rise time $t_\mathrm{on}$ and the domain growth rate $K$, where the rise time is defined as the time taken by the $D(t) - t$ characteristic to change from the 10\% value to the 90\% value. } 
	\label {fig:ton_k} 
\end {figure}

The highest correlation between each parameter was the relationship between $t_\mathrm{on}$ and $K$. The relationship is shown in Fig. \ref{fig:ton_k}. The sample correlation coefficient $\gamma$ is -0.93, and the faster the growth rate, the shorter the time it takes for the total infections to reach $D_\mathrm{s}$. This corresponds to the physical process in which the higher the domain growth rate, the faster the domains will collide with each other and reach the growth limit boundary. 

On the other hand, the value of $\gamma$ between $K$ and $D_\mathrm{s}$ was as small as -0.17. It is suggested that the domain growth rate does not significantly affect the value of $D_\mathrm{s}$. In addition, when $K$ is high and $D_\mathrm{s}$ is also large, the daily infections show a rapid increase and a rapid decrease, and the value of $t_\mathrm{on}$ is shortened.

Since the nucleation decay constant $\nu$ is proportional to the nucleation rate, the correlation with other parameters was similar to that of $K$. The value of $\gamma$ between $t_\mathrm{on}$ and $\nu$ was -0.74 and that between $\nu$ and $D_\mathrm{s}$ was -0.31.

Since the correlation between $D_\mathrm{s}$ and $K$ and the correlation between $D_\mathrm{s}$ and $\nu$ are small, it is necessary to consider the overall infectious nucleation and its growth process that occurs in Japanese society from a broader perspective.

\subsection {Result of the first wave analysis} 

\begin {figure} 
	\centering 
	\includegraphics[width=8.0cm] {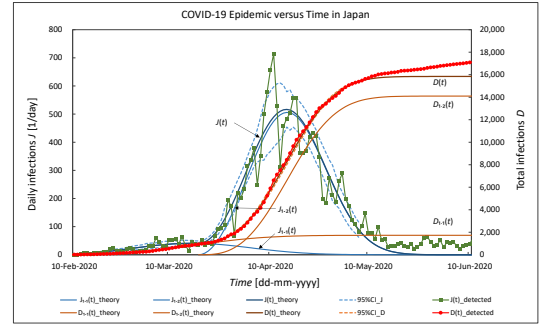} 
	\caption {Detected and simulated results of the first COVID-19 infection wave in Japan. 
	The meanings of the markers and the curves are the same as those in Fig. \ref{fig:allwaves_linear}.
	} 
	\label {fig:firstwave} 
\end {figure}

The first wave was expressed as the superposition of the 1.1th wave and the 1.2th wave. Figure \ref{fig:firstwave} shows the analysis result of the entire first wave. The meanings of the markers and the curves are the same as those in Fig. \ref{fig:allwaves_linear}. The accuracy of the simulated $D$ for the main wave (1.2th wave) was as good as $95\%CI / D < 4\%$.

\subsection {Result of the second wave analysis} 

The theoretical value of daily infections in the 1.2th wave reached a peak of about 506 around April 15, 2020, and decreased to about 30 around May 13, 2020. However, it did not reach zero and increased significantly from around 24th of June\cite{ministry_hlw}. This process was regarded as the second wave and expressed as the superposition of the 2.1th, the 2.2th, and the 2.3th waves. 

\begin {figure} 
	\centering 
	\includegraphics[width=8.0cm] {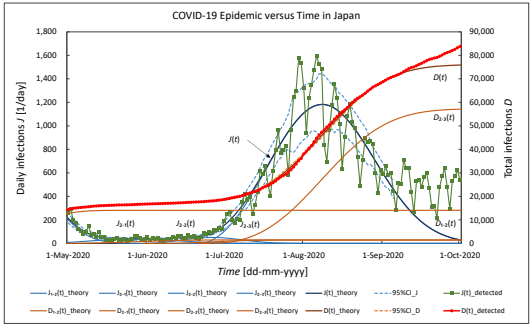} 
	\caption {Detected and simulated results of the second COVID-19 infection wave in Japan. 
	The meanings of the markers and the curves are the same as those in Fig. \ref{fig:allwaves_linear}. 
	}
	\label {fig:secondwave} 
\end {figure}

Figure \ref{fig:secondwave} shows the analysis result of the entire second wave. The meanings of the markers and the curves are the same as those in Fig. \ref{fig:allwaves_linear}. The accuracy of the simulation for the main wave (2.3th wave) was as good as $95\%CI / D < 2.5\%$.

\subsection {Result of the third wave analysis} 

The theoretical value of daily infections in the 2.3th wave reached a peak of about 1,182 around August 8, 2020, and decreased to about 400 around September 7, 2020. However, the number of daily infections did not decrease further and increased significantly in early December\cite{ministry_hlw}. This process was regarded as the third wave and expressed as the superposition of the 3.1th, the 3.2th, and the 3.3th waves. 

\begin {figure} 
	\centering 
	\includegraphics[width=8.0cm] {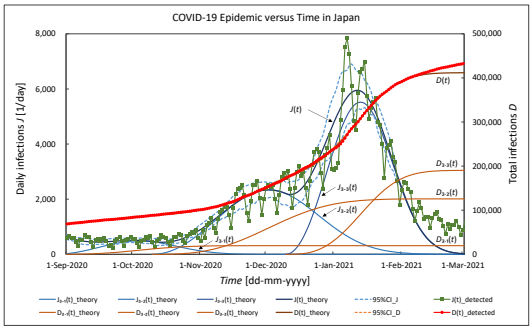} 
	\caption {Detected and simulated results of the third COVID-19 infection wave in Japan. 
	The meanings of the markers and the curves are the same as those in Fig. \ref{fig:allwaves_linear}.
	}
	\label {fig:thirdwave} 
\end {figure}

Figure \ref{fig:thirdwave} shows the analysis result of the entire third wave. The meanings of the markers and the curves are the same as those in Fig. \ref{fig:allwaves_linear}. The accuracy of the simulation for the main wave (3.3th wave) was as good as $95\%CI / D < 1.1\%$.

\subsection {Result of the fourth wave analysis} 

The theoretical value of daily infections in the 3.3th wave reached a peak of about 5,521 around January 13, 2021, and decreased to about 1,570 around February 5, 2021. However, the number of daily infections slowed down since then and increased significantly in early March\cite{ministry_hlw}. This process was regarded as the fourth wave and expressed as the superposition of the 4.1th, the 4.2th, and the 4.3th waves. 

\begin {figure} 
	\centering 
	\includegraphics[width=8.0cm] {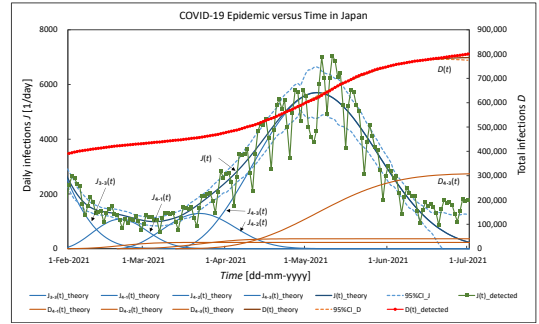} 
	\caption {Detected and simulated results of the fourth COVID-19 infection wave in Japan. 
	The meanings of the markers and the curves are the same as those in Fig. \ref{fig:allwaves_linear}.
	}
	\label {fig:fourthwave} 
\end {figure}

Figure \ref{fig:fourthwave} shows the analysis result of the entire fourth wave. The meanings of the markers and the curves are the same as those in Fig. \ref{fig:allwaves_linear}. The accuracy of the simulation for the main wave (4.3th wave) was as good as $95\%CI / D < 0.5\%$.

\subsection {Result of the fifth wave analysis} 

The theoretical value of daily infections in the 4.3th wave reached a peak of about 5,700 around May 5, 2021, and decreased to about 900 around June 18, 2021. However, the number of daily infections slowed down since then and increased significantly in mid-July\cite{ministry_hlw}. This process was regarded as the fifth wave and expressed as the superposition of the 5.1th, the 5.2th, and the 5.3th waves. 

\begin {figure} 
	\centering 
	\includegraphics[width=8.0cm] {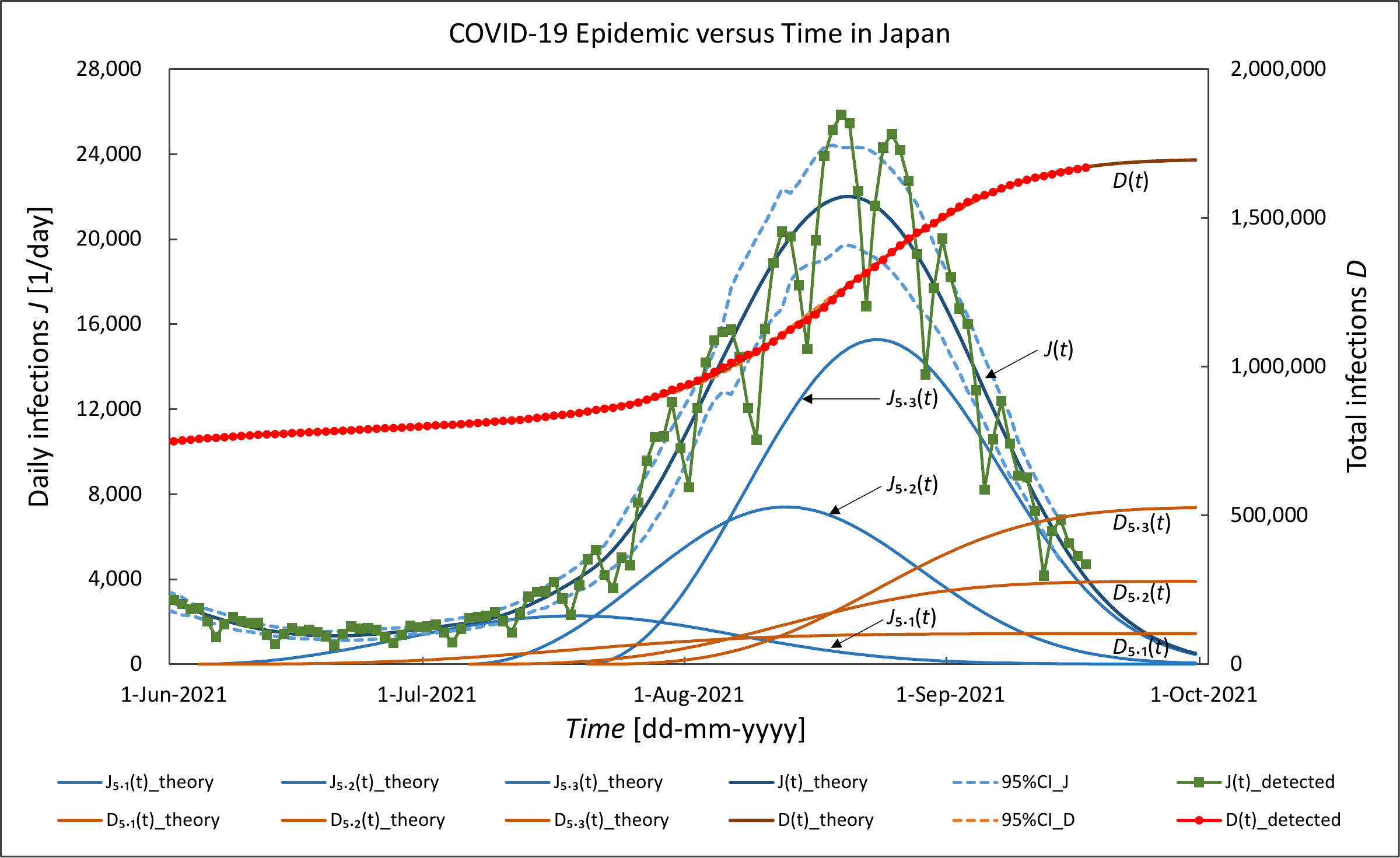} 
	\caption {Detected and simulated results of the fifth COVID-19 infection wave in Japan. 
	The meanings of the markers and the curves are the same as those in Fig. \ref{fig:allwaves_linear}.
	}
	\label {fig:fifthwave} 
\end {figure}

Figure \ref{fig:fifthwave} shows the analysis result of the entire fifth wave. The meanings of the markers and the curves are the same as those in Fig. \ref{fig:allwaves_linear}. The accuracy of the simulation for the main wave (5.3th wave) was as good as $95\%CI / D < 0.7\%$.

\section{Discussions} 
\subsection{Nucleation and growth model} 

In order for one new phase to be produced in the parent phase, a new phase must be born (nucleation) and it must grow\cite{leo_mandelkern}. In the case of COVID-19, the nucleation means that new infections occur among susceptibility holders. However, unlike polarization reversal of ferroelectric materials\cite{ferro_pol_rev, nylon_pol_rev}, new infections do not occur spontaneously. As an entity model, the following physical process is assumed.

Primary infection occurs when infectious virus carriers move into a susceptibility holder population. This primary infection is considered as the nucleation. The time and location of nucleation change randomly as the virus carrier moves. If a new infection occurs at each time and place, it will spread one-dimensionally or two-dimensionally.

\subsection{Infection transmission route seen in the first wave of Japan} 

The SARS-CoV-2 virus which was first confirmed to be infected in Japan in January 2020\cite{ministry_hlw} descended from Wuhan-Hu-1-related isolates\cite{wuhanhu1}, but from February to April 2020, it replaced the D614G variant virus and caused the first wave in Japan. It is said that this variant virus may have been brought from overseas by students who went on a graduation trip in March 2020. Furthermore, March 20 - 22, 2020 were consecutive holidays in Japan, and people's exchanges increased. Then, in early April, the number of infections increased rapidly, forming the first wave\cite{genomeanalysis,firstwave}.

\begin {figure} 
	\centering 
	\includegraphics [width=8.0cm] {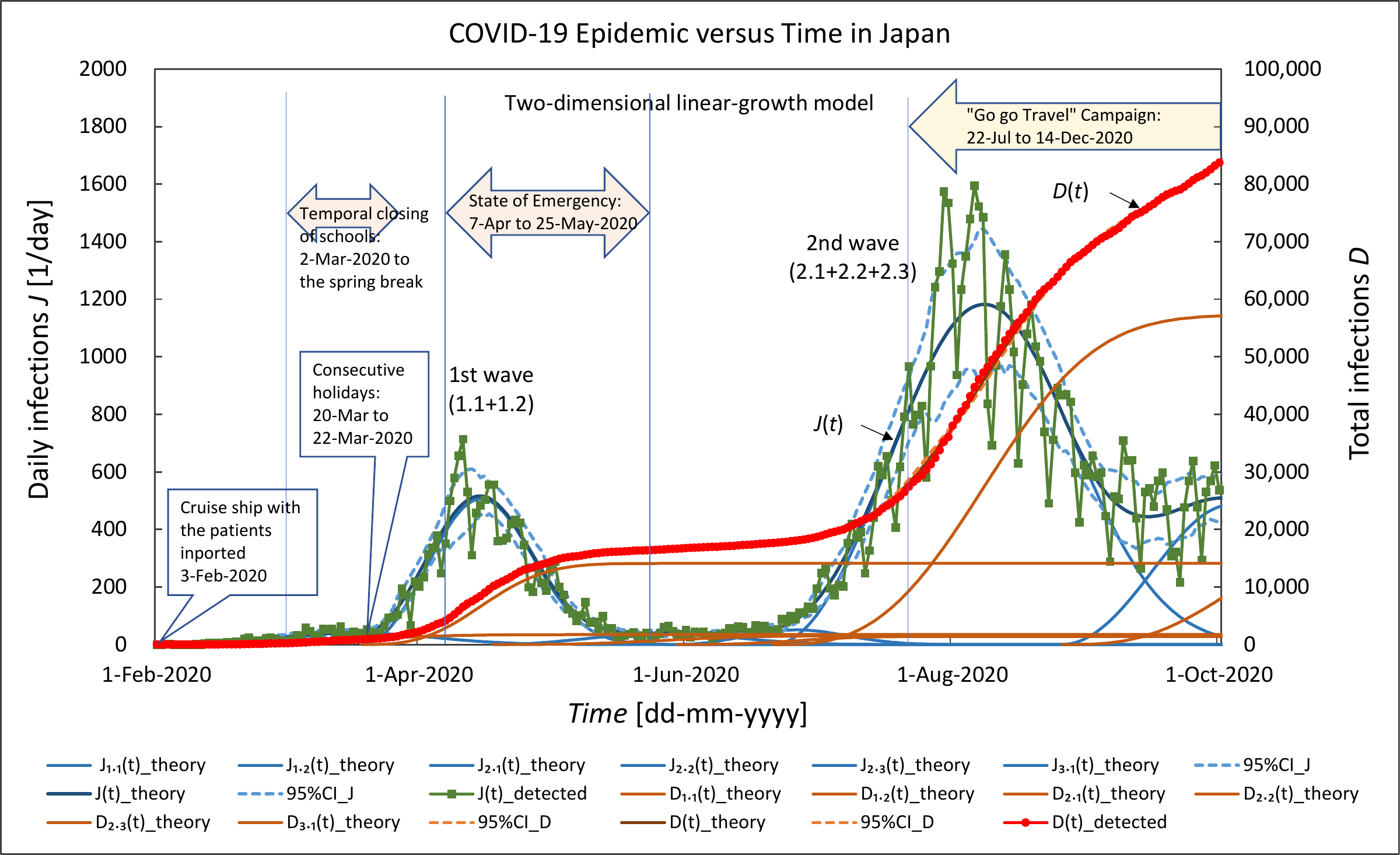} 
	\caption {Detected and simulated results of the first and the second COVID-19 infection waves in Japan. 
	The meanings of the markers and the curves are the same as those in Fig. \ref{fig:allwaves_linear}.
	}
	\label {fig:1st_2ndwaves} 
\end {figure}

Figure \ref{fig:1st_2ndwaves} shows the characteristics of the first and the second waves. The meanings of the markers and the curves are the same as those in Fig. \ref{fig:allwaves_linear}. The government issued a state of emergency on April 7, 2020 in seven prefectures including Tokyo prefecture, and expanded it nationwide on 16th of April\cite{stateofemergency1}. The daily number of infections in the first wave reached a maximum of 714 on April 11, 2020, then declined to 28 on May 25 when the state of emergency was lifted nationwide. As can be seen from Fig. \ref{fig:1st_2ndwaves}, the number of new infections is low during June, but increases again in early July, forming a second wave larger than the first wave, though it is necessary to consider that the PCR test was performed in a limited manner during the first wave.

The government did not declare a state of emergency during the second wave. However, the inspection system was also in place from the time of the first wave, and it became possible to grasp the whole picture of infected people and people refrained from acting, so the second wave reached maximum on August 6, 2020\cite{2ndwave_kutuna}. After reaching 1,595, it decreased along the theoretical curve, as can be seen in Fig. \ref{fig:1st_2ndwaves}.

\subsection{Economic measures and consecutive holiday effect}

\begin {figure} 
	\centering 
	\includegraphics [width=8.0cm] {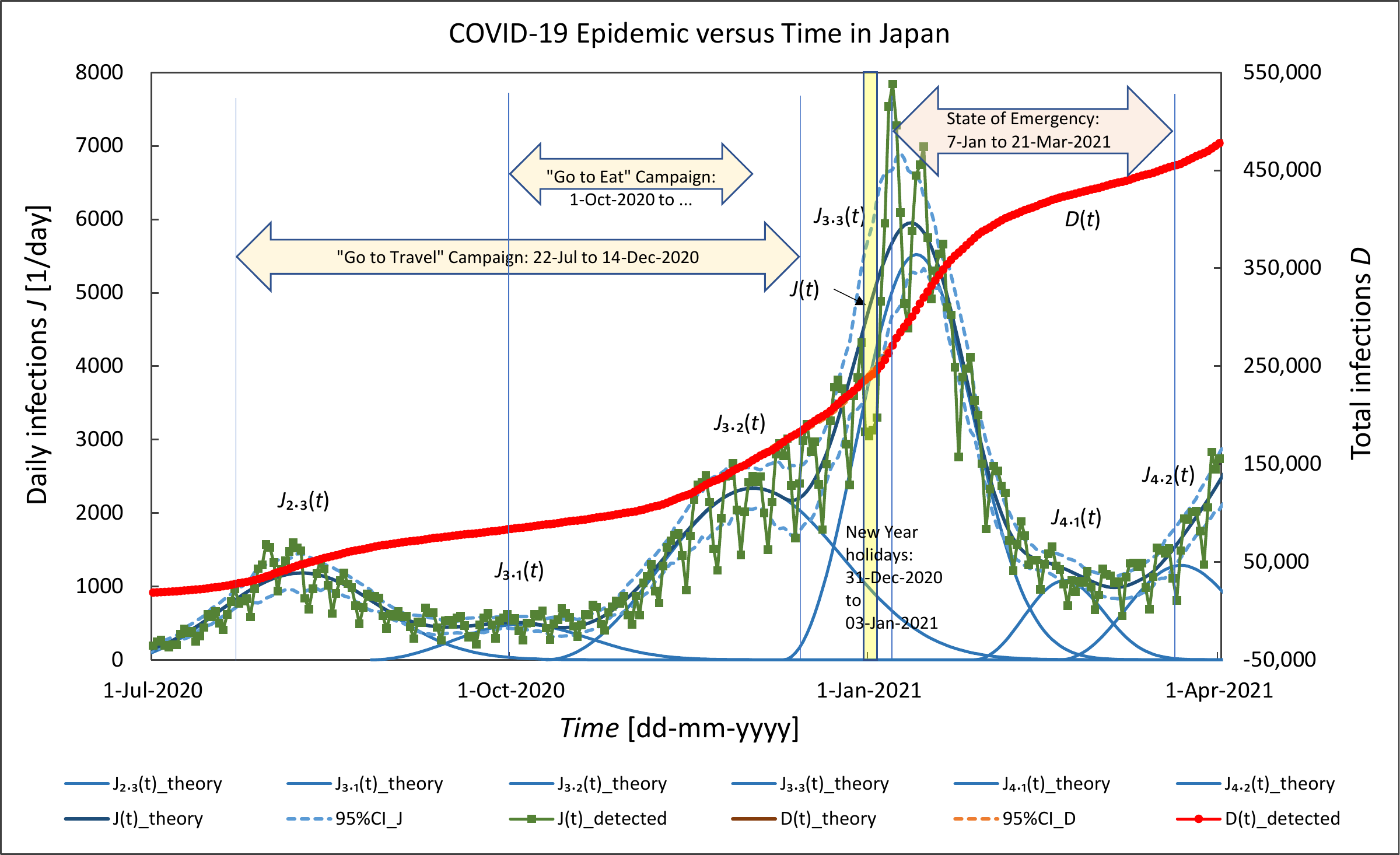} 
	\caption {Detected and simulated results of the second and the third COVID-19 infection waves in Japan. 
	The meanings of the markers and the curves are the same as those in Fig. \ref{fig:allwaves_linear}.
	}
	\label {fig:2nd_3rdwaves} 
\end {figure}

Figure \ref{fig:2nd_3rdwaves} shows the characteristics of the second and the third waves of COVID-19 in Japan. The figure shows the periods of "Go to Travel Campaign" (from 22-Jul-2020 to 14-Dec-2020) and "Go to Eat Campaign" (from 1-Oct-2020 to …) projects carried out by the government of Japan\cite{goto_campaign}. 

The "Go to Travel Campaign" was held during the second wave, but the impact on the infection was ignored because the second wave decreased in September as seen in Fig. \ref{fig:1st_2ndwaves}. However, in November, when the two campaigns were underway, the 3.2th wave, which was larger than the second wave, occurred. For this reason, the government announced a nationwide suspension of the projects. Immediately after the announcement of this suspension, the third wave increased more rapidly and formed the largest ever 3.3th wave as seen in Fig. \ref{fig:2nd_3rdwaves}.

The consecutive holidays from December 31, 2020 to January 3, 2021 are New Year holidays, and as a tradition, many Japanese people move and interact nationwide.

\begin {figure} 
	\centering 
	\includegraphics [width=8.0cm] {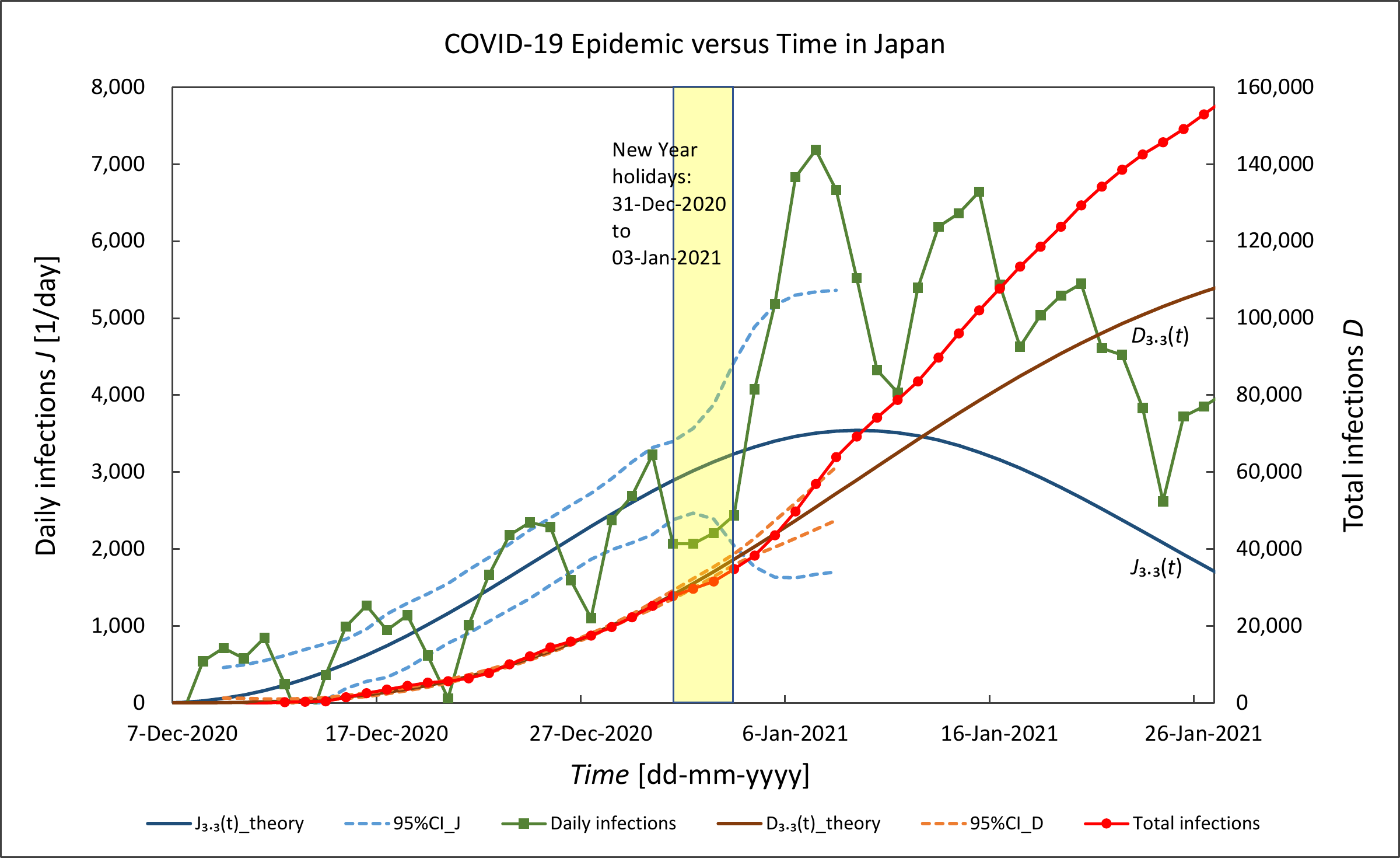} 
	\caption {Detected values and simulated results before the holidays of the 3.3th COVID-19 infection wave in Japan.
	The meanings of the markers and the curves are the same as those in Fig. \ref{fig:allwaves_linear}.
	}
	\label {fig:3rd_3wave} 
\end {figure}

Figure \ref{fig:3rd_3wave} shows the characteristics of extracted 3.3th wave from the third wave. The thick blue and the brown curves represent the theory of the daily and the total infections, respectively, obtained from the detected infections before the holidays (until December 31, 2020). 

When there are consecutive holidays, the detected value of the daily infections before the holidays fluctuates regularly in a weekly cycle, and the data in the holidays shows irregularly small values. Then, the data returns to the one-week cycle type after showing irregularly large values. The data in the holidays is small because many medical institutions are also closed. It is clear that the number of infections after the holidays increased significantly compared with the simulation before the holidays. 

\begin {figure} 
	\centering 
	\includegraphics [width=8.0cm] {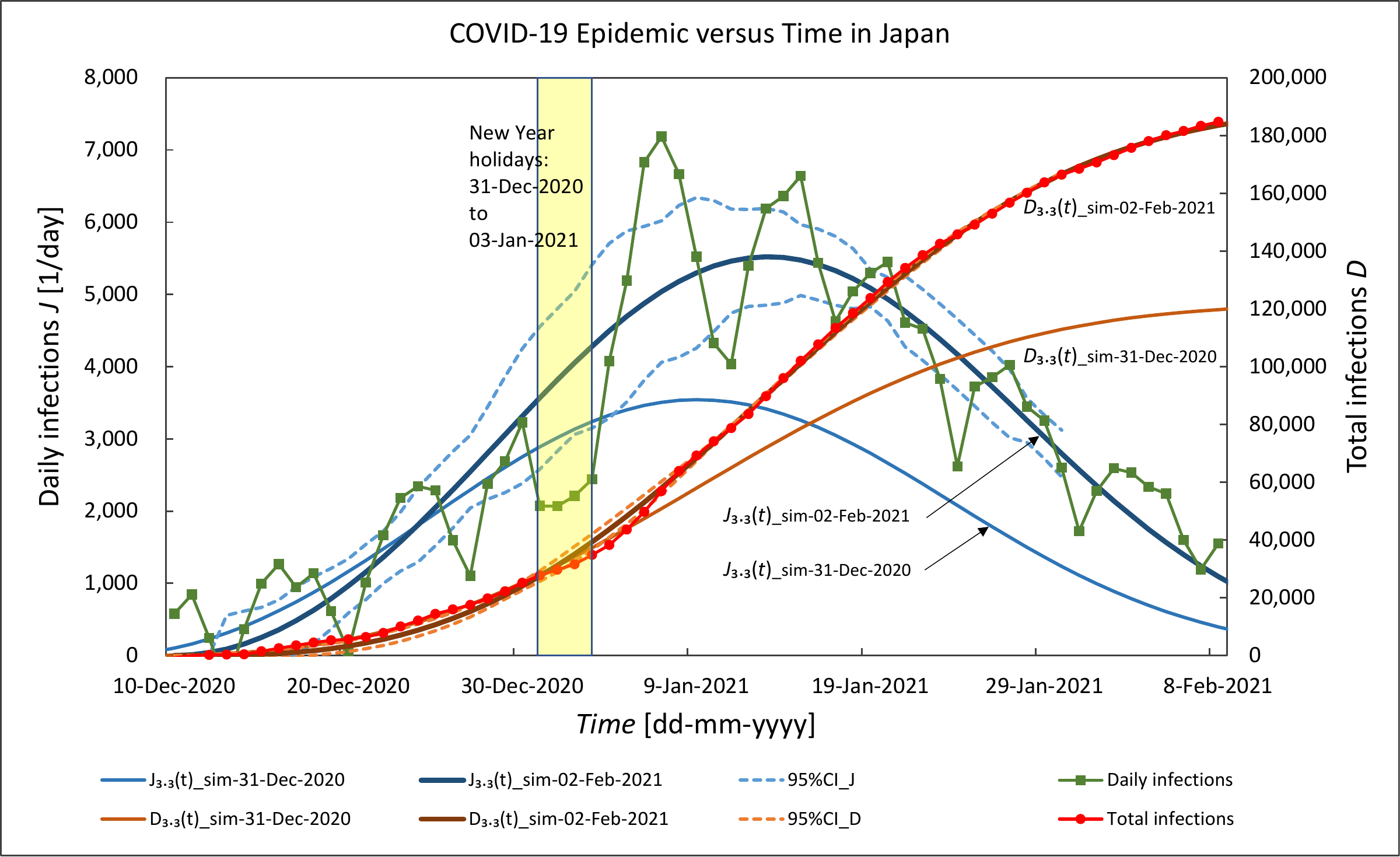} 
	\caption {Detected and two simulated results before and after the holidays of the 3.3th COVID-19 infection wave in Japan.
	The meanings of the markers and the curves are the same as those in Fig. \ref{fig:allwaves_linear}.
	}
	\label {fig:3rd_3wave_comp} 
\end {figure}

Figure \ref{fig:3rd_3wave_comp} shows the characteristics of the extracted 3.3th wave showing two simulated curves.
The thick blue and the brown curves represent the theory of the daily and the total infections, respectively, obtained from the detected infections after the holidays (until February 3, 2021). 
The thin blue and the brown curves represent the theory of the daily and the total infections, respectively, which are the same ones as in Fig. \ref{fig:3rd_3wave}.

The characteristic of the 3.3th wave is that the speed of infection spread is about 1.6 times faster than that of the 3.2th wave (from Table \ref{table:three_parameters}, $\sqrt{0.0106} / \sqrt{0.0043} \simeq 1.6$) and $D_\mathrm{s}$ is also large. The daily infections of the 3.3th wave have shown a rapid increase and a rapid decrease. This is one of the common characteristics mentioned in the "Determined parameters" subsection. Therefore, this is the natural infectious properties of the virus in the lifestyle of Japanese people including the effect of the state of emergency that was issued by the government of Japan on January 7, 2021\cite{stateofemergency2}. 

It is reasonable to consider that during the period of the two economic measures, the movement and interaction of people increased nationwide leading the infection to the large 3.2th wave, and the New Year holidays accelerated the infection more rapidly leading to the largest ever 3.3th wave.

\begin {figure} 
	\centering 
	\includegraphics [width=8.0cm] {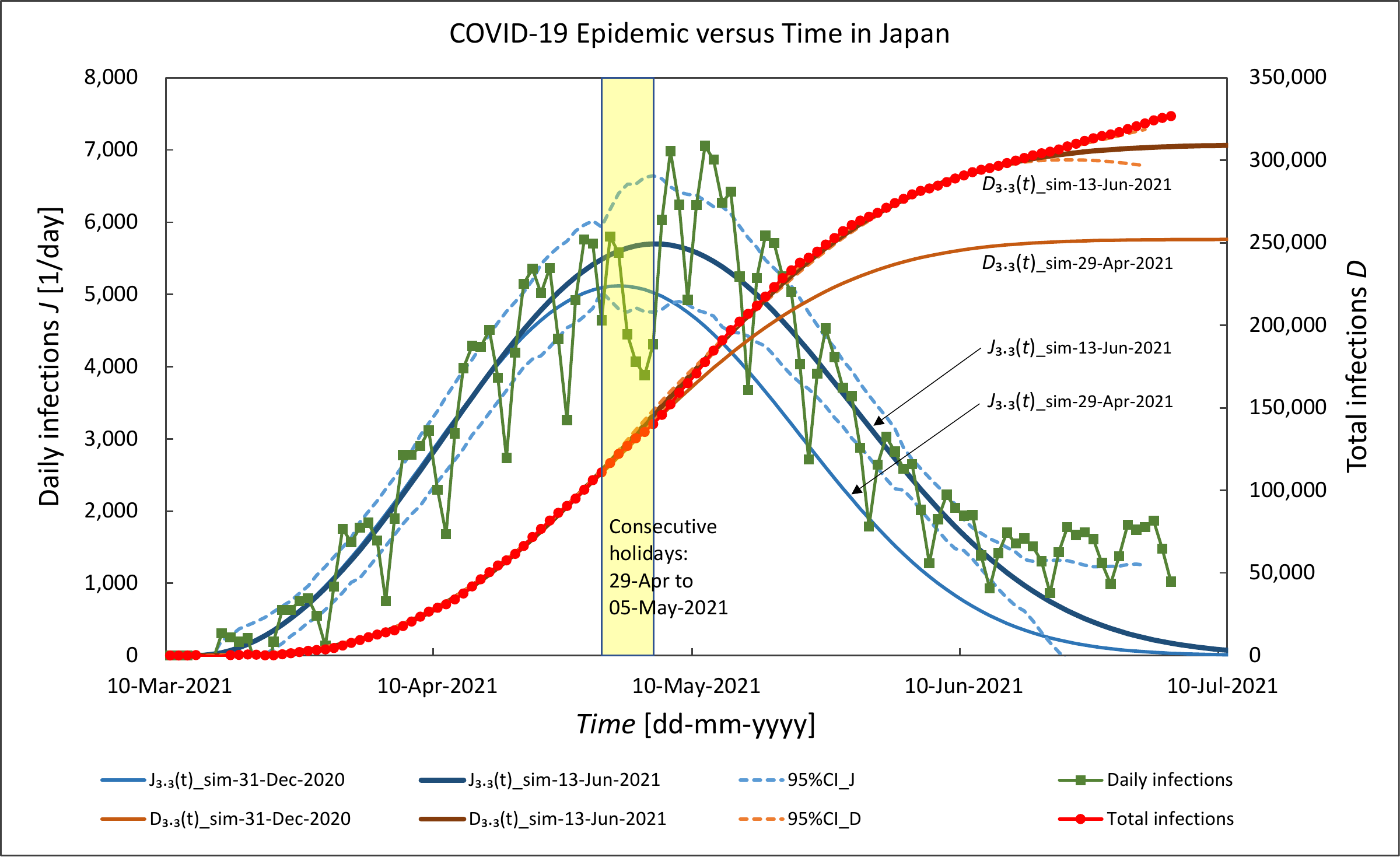} 
	\caption {Detected and two simulated results before and after the holidays of the 4.3th COVID-19 infection wave in Japan.
	The meanings of the markers and the curves are the same as those in Fig. \ref{fig:allwaves_linear}.
	}
	\label {fig:4th_3wave_comp} 
\end {figure}

Another traditional holiday in Japan is called "Golden Week", the consecutive holidays in early May. Figure \ref {fig:4th_3wave_comp} shows the 4.3th wave characteristics of analyzing "Golden Week" from April 29 to May 3, 2021 as the same way as in Fig. \ref{fig:3rd_3wave_comp}.

Table \ref{table:predictability} shows the results of the analysis carried out to know the predictability of the number of infections in the near future using the characteristics of the 4.3th wave in addition to the analysis of consecutive holidays. If the analysis duration of the 4.3th wave in Table \ref {table:waves_duration} is from March 13, 2021 to 15th of May, 28th of May, and 13th of June, the ratio of the former two values of $D_\mathrm{s}$ to the last $D_\mathrm{s}$ were 99.6\% and 101.7\%, respectively. The number of infections in the near future could be predicted within $\pm 2\%$ from the detected values after the number of daily infections exceeded the maximum value where the environmental change was small.

\begin{table} 
  \caption {The simulation predictability} 
  \label {table:predictability} 
  \centering 
  \begin {tabular} {crr} 
	\hline 
	End day & \multicolumn{1}{c}{$D_\mathrm{s}$} & \multicolumn{1}{c}{$D_\mathrm{s}$ ratio}\\ 
	of simulation & [person] & \multicolumn{1}{c}{[$\si{\%}$]} \\ 
	\hline \hline 
	15-May-2021 & 308,253 & 99.6 \\ 
	28-May-2021 & 314,596 & 101.7 \\  
	13-Jun-2021 & 309,360 & 100.0 \\ 
	\hline
  \end {tabular} 
\end {table}

Table \ref{table:holidayeffect} shows the simulated values of $D_\mathrm{s}$ before and after the holidays for the two "consecutive holiday effect". This time, the “New Year holiday effect” was 156\% and the "Golden Week holiday effect" was 123\%.

\begin {table} 
\caption {The values of $D_\mathrm{s}$ of New Year holidays and Golden Week holidays simulated before and after the holidays.}
\label {table:holidayeffect} 
\centering
  \begin {tabular} {ccccc} 
	\hline 
	Holiday & $D_\mathrm{s-bef}$ & $D_\mathrm{s-aft}$ & Effect\\ 
	 & [person] & [person] & [\%] \\ 
	\hline \hline 
	New Year & 121,966 & 190,417 & 156 \\ 
	Golden Week & 252,099 & 309,513 & 123 \\ 
	\hline 
  \end {tabular} 
\end {table}

\section {Conclusion}
The purpose of this study was to simulate all COVID-19 infection waves that occurred repeatedly from February 2020 to September 2021 in Japan. The entire waveform was basically simulated by the superposition of five major waves, and the least-squares method was applied to each wave to determine the parameters of the Avrami equation. Despite the simulation using the single equation assuming that the parameters were constants, the accuracy of the simulated cumulative infection $D$ was as good as $95\% CI / D < 2.5\%$ after the second wave when $D$ exceeded 10,000. Since the simulation was highly accurate, it was effective in predicting the number of infections in the near future and in considering the cause of changes in the number of infections observed as the detailed structure of the waveform. On the other hand, the relevance of this method to the conventional SIR model has not been investigated.

\end{document}